\documentclass[twocolumn,amsmath,amssymb,prb,superscriptaddress,longbibliography]{revtex4-1}
\usepackage{graphicx}
\draft

\begin{document}

\preprint{}

\title{Deterministic Writing and Control of the Dark Exciton Spin using
Short Single Optical Pulses} 

\author{I. Schwartz}
\thanks{These authors contributed equally to this work.}
\author{E. R. Schmidgall}
\thanks{These authors contributed equally to this work.}
\author{L. Gantz}
\affiliation{The Physics Department and the Solid State Institute,
Technion--Israel Institute of Technology, 32000 Haifa, Israel}
\author{D. Cogan}
\affiliation{The Physics Department and the Solid State Institute,
Technion--Israel Institute of Technology, 32000 Haifa, Israel}
\author{E. Bordo}
\affiliation{The Physics Department and the Solid State Institute,
Technion--Israel Institute of Technology, 32000 Haifa, Israel}
\author{Y. Don}
\affiliation{The Physics Department and the Solid State Institute,
Technion--Israel Institute of Technology, 32000 Haifa, Israel}
\author{M. Zielinski}
\affiliation{Institute of Physics, Faculty of Physics, Astronomy and Informatics, Nicolaus Copernicus University, ul
Grudziadzka 5, PL-87-100 Torun, Poland}
\author{D. Gershoni}
\affiliation{The Physics Department and the Solid State Institute,
Technion--Israel Institute of Technology, 32000 Haifa, Israel}
\email[]{dg@physics.technion.ac.il}

\date{\today}

\begin{abstract}
We demonstrate that the quantum dot-confined dark exciton forms a long-lived
integer spin solid state qubit which can be deterministically on-demand initiated in a
pure state by one optical pulse. Moreover, we show that this qubit can be
fully controlled using short optical pulses, which are several orders of
magnitude shorter than the life and coherence times of the qubit. Our
demonstrations do  not require an externally applied magnetic field, and they
establish that the quantum dot-confined dark exciton forms an excellent
solid state matter qubit with some advantages over the half-integer spin
qubits, such as the confined electron and hole, separately. Since quantum dots
are semiconductor nanostructures that allow integration of
electronic and photonic components, the dark exciton may have important
implications for implementations of quantum technologies consisting of
semiconductor qubits.
\end{abstract}

\pacs{xyz.123, xyz.123}

\maketitle
\section{Introduction}
The ability to coherently control and exploit matter quantum systems is
essential for realizations of future technologies based on quantum information processing (QIP).\cite{divincenzo2000, kimble2008} Optical approaches, in
particular, are preferred since they are state-selective, require no
contacts, and are ultrafast.\cite{hennessy2007}  Semiconductors play an
important role among the matter venues of choice, since they dovetail with
contemporary leading technologies in general, and those of light sources and
detectors in particular. For these reasons, optical control of spins of
charge carriers in semiconductor quantum dots (QDs) has been the subject of
many recent works.\cite{obrien2009, kim2010, press2010, ramsay2008,
greilich2011, degreve2011, godden2012} While self-assembled QDs leave yet unresolved the issues of scalability required for QIP due to the stochastic nature of their size, shape, and composition, these systems do provide a convenient platform for extensive investigations of single- and few-carrier matter qubit systems with extremely efficient coupling to light.

The absorption of a photon in semiconductors promotes an electron from the full valence band to the empty conduction band. This fundamental excitation is particularly efficient because the valence band is formed from molecular $p$-like electronic orbitals, while the conduction band is formed from $s$-like orbitals, and the dipole moment between these orbitals strongly interacts with the electric field of the light. The photo-excitation, thus, does not alter the promoted electron spin. The excitation leaves an excited electron in the conduction band and a missing one in the valence band. This many-electron configuration can be described as a two-body state if the missing electron is treated as a hole with opposite quantum numbers (positive rather than negative charge and spin-up rather than spin-down).~\cite{cardona} In this description, an absorbed photon results in an electron-hole (e-h) pair with antiparallel spins, or a bright exciton (BE). In InAs/GaAs self assembled QDs, the quantum size and the lattice mismatch strain result in a considerable energy difference between holes which have their spin direction aligned with their orbital molecular momentum (heavy holes: total angular momentum projection of $\pm 3/2$ on the QD growth direction) and holes with a spin direction which is antiparallel to the orbital momentum (light holes: total angular projection of $\pm 1/2$). The lowest energy BEs are thereby composed of an electron-heavy hole pair with total integer spin projections of $\pm 1$ on the QD growth axis, reflecting the difference in orbital momentum between the ground (valence) and excited (conduction) electron states. Since the orbital momentum is aligned with the electronic spin, the polarization of light can be straightforwardly used to coherently ``write," ``read," and manipulate the spin of BEs in self assembled QDs.\cite{benny2011prl, poem2011, kodriano2012,muller2013}

BEs have many advantages over single half-integer carrier spins, for which deterministic writing and coherent control by a single ultrashort pulse is impossible. Excitons are also advantageous in their electric neutrality,
making them less susceptible to decoherence due to electrostatic
fluctuations in their vicinity.\cite{kim2010, press2010, ramsay2008,
greilich2011, degreve2011} In addition, their heavy hole content partially protects
them from decoherence due to nuclear magnetic field fluctuations.\cite{fischer2008,gerardot2008,
greilich2011, degreve2011, godden2012} Their typically short radiative
lifetimes (about 1 nsec), however, limit their usefulness.

A dark exciton (DE) is an electronic excitation in which the hole spin is parallel to that of the electron.
In QDs, the DE has total integer spin of 2,\cite{poem2010} with
projections of $\pm2$ on the QD growth axis, reflecting the difference in both
angular momentum and spin between the conduction and valence band electron states.
Since photons barely interact with electronic spin, the DE is almost
optically inactive and has a lifetime that is orders of magnitude longer
than that of the BE.\cite{mcfarlane2009}

In this work, we shed new light on the residual optical activity of the DE and directly measure its radiative lifetime. We then circumvent its very weak optical activity and experimentally demonstrate that the QD-confined dark exciton forms a novel semiconductor matter qubit system. We deterministically generate it in one eigenstate using one single optical pulse.
We obtain a lower bound of about a hundred nsecs on its coherence time
and use resonant optical pulses to coherently control its spin.

\section{Experimental System}

The measurements were carried out on a single strain-induced InGaAs QD embedded within a one-wavelength planar microcavity.
The microcavity design facilitates efficient collection of the light emitted from single QDs
whose optical transitions resonate with the microcavity mode.~\cite{ramon2006}
The sample was grown by molecular beam epitaxy on a [001]-oriented GaAs substrate.
One layer of strain-induced InGaAs QDs was deposited in the center of a 285-nm-thick intrinsic GaAs layer.
The GaAs layer was placed between two distributed Bragg reflecting mirrors consisting of
25 (bottom) and 10 (top) periods of pairs of AlAs/GaAs quarter-wavelength-thick layers.
For the optical measurements, the sample was placed inside a sealed metal tube
cooled by a dry helium refrigerator maintaining a temperature of 4.0K.
A $\times$60 microscope objective with numerical aperture of 0.85 was placed above the sample
and used to focus the light beams on the sample surface and to collect the emitted photoluminescence (PL).
Continuous wave and/or pulsed laser excitations were used.
Acousto-optic modulators were used to produce pulses of few tens of nanoseconds duration.
Picosecond pulses were generated by synchronously pumped, cavity-dumped dye lasers.
The temporal width of the dye laser pulses was $\sim$10 psec and their spectral width was $\sim$100 $\mu$eV.
The polarizations of the pulses were independently adjusted using a polarized beam splitter (PBS)
and two pairs of computer-controlled liquid crystal variable retarders (LCVRs).
The polarization of the emitted PL was analyzed by the same LCVRs and PBS.
The PL was spectrally analyzed by 1-meter monochromators and detected by either silicon avalanche
photodetectors coupled to a HydraHarp 400$^{TM}$ time-correlated single photon counter or by a
cooled charged coupled array detector.~\cite{benny2011prb}

\section{Results}
\subsection{Observation of Photoluminescence from the Dark Exciton}

\begin{figure}
\begin{center}
\includegraphics[width=\columnwidth]{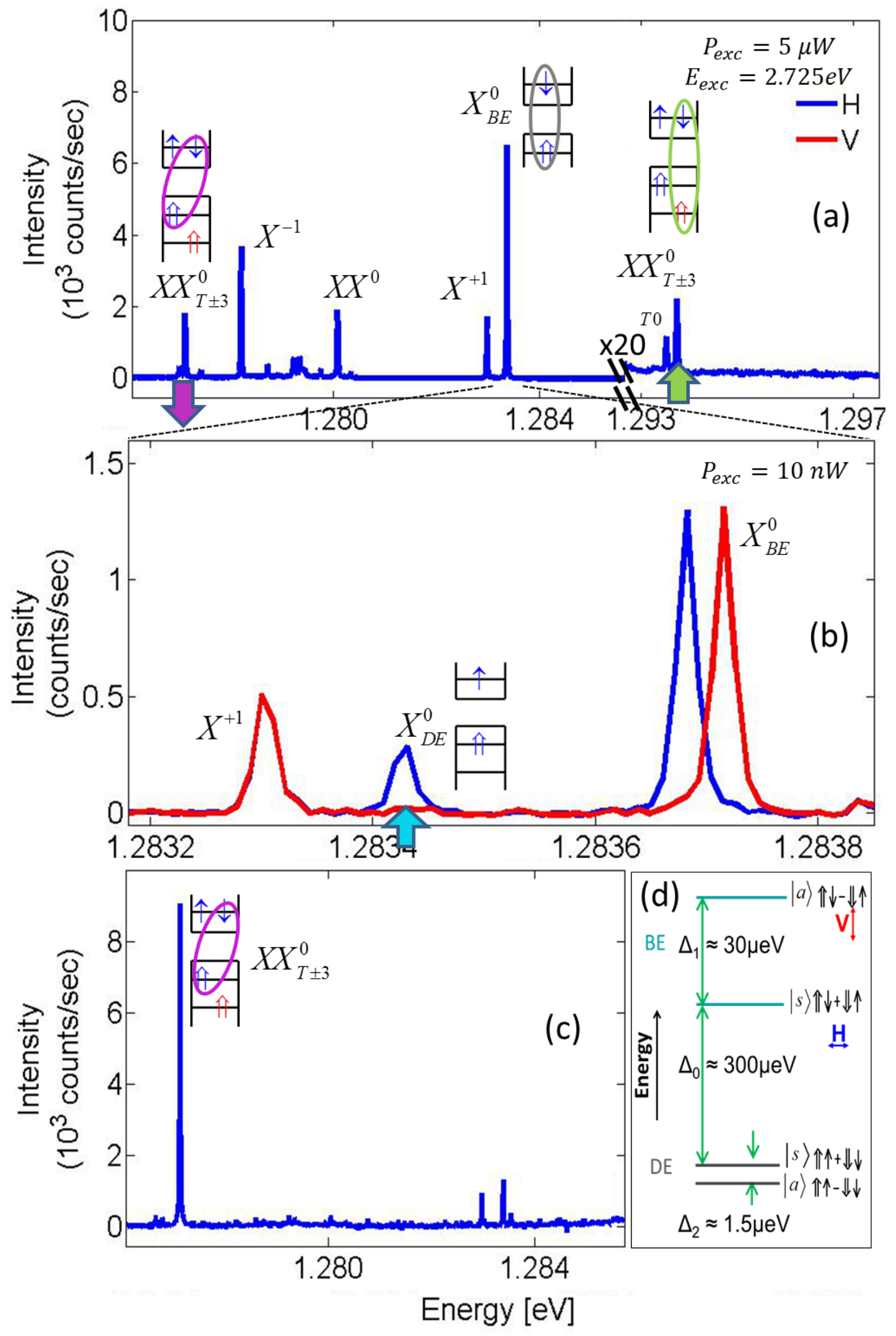}
\caption{Photoluminescence spectra of the single QD -observation of light emission from the DE. (a) PL spectrum of the QD used in this research excited by high intensity (5 $\mu$W) cw light. The lines are named by their initial states from which the radiative recombination originates. Above the most relevant lines, carrier configurations are schematically displayed. $\uparrow$ ($\Uparrow$) denotes the electron (heavy hole) spin state and a short blue (red) arrow denotes $s$- ($p$-) shell single carrier level. Only one of the two possible spin configurations is displayed in each case. The recombining e-h pair is indicated by a matching oval. The spectral line used for detecting (exciting) the $XX_{T\pm3}^{0}$  biexciton line is marked by downward magenta (upward green) arrow. (b) Polarization-sensitive, expanded scale PL spectrum of the QD at low excitation intensity (10 nW). At this excitation level, the DE spectral line is clearly observed about 300 $\mu$eV below the BE lines. Unlike the BE line, which has two cross rectilinearly polarized components, the DE has one horizontal polarization.  (c) PL spectrum from the QD under resonant excitation into the higher energy optical transition of the  $XX_{T\pm3}^{0}$  biexciton line (upward green arrow in (a)). (d) Schematic description of the energy fine structure of the QD confined exciton.}
\label{fig:spectrum}
\end{center}
\end{figure}

Figure \ref{fig:spectrum}(a) presents a PL spectrum of a single QD excited non-resonantly by a 445nm (2.78 eV)
continuous wave laser. Excitation at such a high energy above the bandgap photogenerates electron-hole pairs in the bulk semiconductor. The photogeneration rate corresponds to the intensity of the exciting laser. Very small numbers of these carriers, uncorrelated and with randomized spin directions,\cite{poem2010} arrive to the QD and populate its lowest energy levels.
In  50\% of the cases e-h pairs with antiparallel spins populate the QD, and BEs are formed. A BE ($X_{BE}^{0}$) will recombine radiatively within $\sim$1 nsec, giving rise to an observed PL spectral line.
Optical recombination of e-h pairs with anti parallel spins from different configurations of charge carriers, such as the charged excitons $X^{+1}$ and $X^{-1}$, results in spectral lines at different energies from different configurations. The identification of these lines has been the subject of substantial research.\cite{poem2010, benny2011prb, kodriano2010, poem2011}

In this work, we are primarily interested in four of the spectral lines visible in Figure \ref{fig:spectrum}.  The initial carrier configurations from which e-h pair recombination gives rise to the observed PL are schematically described above each of these lines. The first emission line ($X_{BE}^{0}$) results from recombination of an electron-hole pair (oval matched in the scheme) in their ground $s$-shell-like QD levels.  The second and third spectral lines initiate from the initial configuration of the spin-blockaded biexciton ($XX_{T\pm3}^{0}$; This configuration has angular momentum projection of $\pm 3$). It is composed of two electron-heavy-hole pairs, where the electrons form a spin singlet in their ground level and the holes form a spin-parallel triplet with one hole in the QD ground level and one in the first excited level. The line at $\sim$1.277eV corresponds to recombination of an electron and a ground level hole (oval matched in the scheme), and the much weaker line (about 50 times weaker, due to the reduced overlap between electron and hole wavefunctions of different symmetries) at $\sim$1.294 eV corresponds to recombination of an electron with the excited hole. In either case, radiative recombination of this biexciton leaves in the QD an electron-hole pair with parallel spins, or a dark exciton (DE).

As demonstrated and discussed below, the DE is not totally dark, and it has residual optical activity
which results in a very slow radiative decay.~\cite{poem2010, mcfarlane2009}
Consequently, at high excitation intensities, the rate of carrier accumulation in the QD exceeds the DE radiative rate by a large margin. Additional carriers destroy the DE, and PL emission from it cannot be observed. In contrast, at very low excitation rates, the carrier accumulation rate is comparable to the DE radiative rate, and the probability of finding the QD occupied with a DE is very significant. Consequently, if the DE recombines radiatively, its PL emission intensity should become comparable to that of the BE.

Indeed, this is what we demonstrate in Figure \ref{fig:spectrum}(b). Here, at a non-resonant excitation intensity which is about three orders of magnitude weaker than that used in Figure \ref{fig:spectrum}(a), the probability of adding an additional carrier to the QD during the DE lifetime is fairly low, and the PL from the DE spectral line ($X_{DE}^{0}$)is clearly observed. This line marked by an upward blue arrow in Figure \ref{fig:spectrum}(b) is the fourth spectral line of interest.

In Figure \ref{fig:spectrum}(b), we present rectilinear polarization-sensitive PL spectra of
the QD at an excitation power that is almost three orders of magnitude lower
(~10nW) than that used for Figure \ref{fig:spectrum}(a). At $\sim$1.2834 eV, approximately 300 $\mu$eV below the
BE lines, a linearly-polarized spectral line is clearly seen. The line
polarization matches that of the lowest energy component of the BE ($H$-
along the major axis of the QD) and its energy difference from the BE
corresponds well to previous measurements of the DE-BE energy separation in
a similar sample\cite{alon2006} and in others.\cite{bayer2002,ortner2003}
Only one linearly polarized transition associated with the DE recombination is observed.
This is in contrast to the BE, which has two almost equally strong cross linearly
polarized transitions.

In Figure \ref{fig:spectrum}(c) we present a PL spectrum from the QD under resonant excitation into the higher energy optical transition of the  $XX_{T\pm3}^{0}$  biexciton line (upward green arrow in Figure \ref{fig:spectrum}(a)) in the presence of very low intensity ($\sim$ 5 nW) non resonant blue laser excitation. Under these conditions, the DE is resonantly excited into the $XX_{T\pm3}^{0}$  biexciton state and the later most efficiently recombines radiatively, as clearly demonstrated by the dominating emission line due to the lower energy optical transition of the biexciton (downward magenta arrow in Figure \ref{fig:spectrum}(a)). The absence or weakness of all other spectral lines indicates that under these conditions of resonant excitation, the QD remains neutral  and it is mainly occupied with DE and its biexciton $XX_{T\pm3}^{0}$.

\begin{figure}
\begin{center}
\includegraphics[width=\columnwidth]{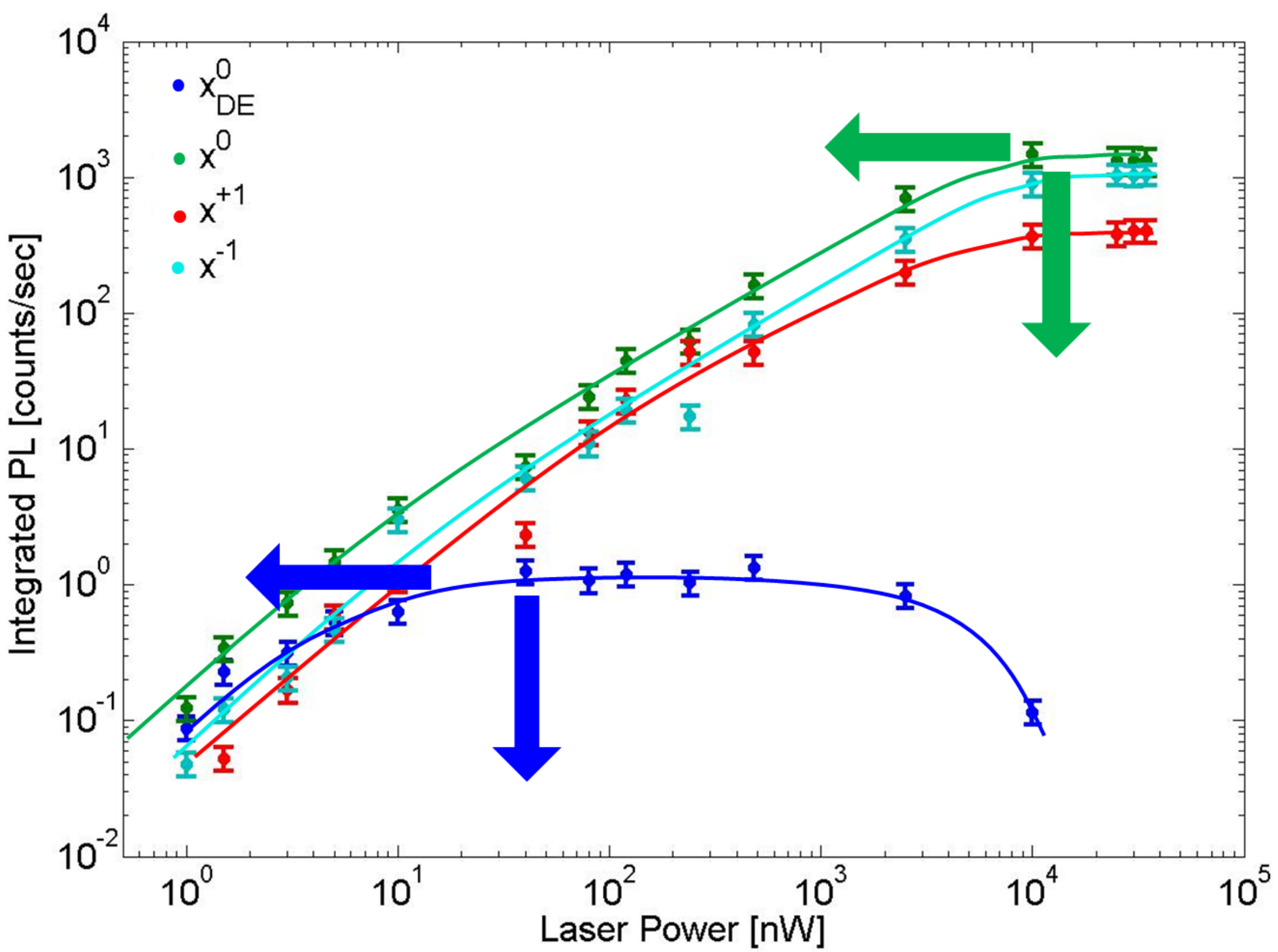}
\caption{Emission intensity from various exciton lines as a function of the laser excitation power. Note that the maximum emission of the DE (blue arrow) is about $\sim$ 3 orders of magnitude weaker than that of the BE (green arrow) and that the power at which the DE maximum occurs is about $\sim$ 3 orders of magnitude weaker than that of the BE. These measurements demonstrate that the DE decays radiatively with oscillator strength which is three orders of magnitude weaker than that of the BE.}
\label{fig:PL_power}
\end{center}
\end{figure}

Figure \ref{fig:PL_power} presents a measurement of the emission intensities of the DE, BE, and the two singly-charged excitons $X^{+1}$ and $X^{-1}$ as a function of the power of the off-resonant excitation. One clearly sees
that the DE emission rate saturates at an excitation power which is three
orders of magnitude lower than the power needed to saturate the emission
from the bright excitons (marked by vertical arrows in Figure \ref{fig:PL_power}). Moreover, we note that the maximum emission intensity from the DE is about
three orders of magnitude weaker than that from the BE (horizontal arrows in Figure \ref{fig:PL_power}). This unambiguously indicates that
the DE lifetime is radiative (like that of the BE) and that it is three orders of magnitude longer
($\sim 1$ $\mu$sec) than that of the BE. We verify this straightforward
conclusion in two additional independent ways: (1) by measuring the
oscillator strength for absorption into the DE line and (2) by
directly measuring the DE lifetime after its deterministic initialization.

\subsection{The Optical Activity of the Dark Exciton}

The residual optical activity of the DE in rotationally symmetric QDs is attributed to light hole/heavy hole mixing. This mixing results in a small, non-vanishing dipole matrix element, polarized along the QD symmetry axis, for only one of the DE eignestates.~\cite{ivchenko2000, dupertuis2011} Realistic atomistic calculations for symmetric InGaAs/GaAs QDs estimate that this DE dipole moment is 5-6 orders of magnitude weaker than that of the BE.~\cite{smolenski2012, korkusinski2013,zielensky2013}

In contrast, we experimentally showed in Figure \ref{fig:spectrum}(b) a much stronger (only 3 orders of magnitude weaker than the BE) in-plane polarized optical activity of one of the DE eigenstates.
The QD that we report on here (and a few similar ones we measured) is not that symmetric.
In reality, self assembled semiconductor QDs are never ideally symmetric. Their deviation from the combined structural and lattice $C_{2v}$ symmetry (two-fold rotations around the symmetry axis and two reflections about two perpendicular mirror planes which include the symmetry axis) can be due to anisotropic composition and/or strain distribution~\cite{bayer2002}  or due to the formation of facets on the interface between the QD and the host material.~\cite{zielinsky2014}
The influence of this symmetry reduction on the confined DE and BE has been quantitatively studied using atomistic model calculations.~\cite{zielinsky2014}
In short, in a QD with reduced symmetry
BE/DE mixing is anticipated.~\cite{bayer2002}
Only a relatively small such mixing is required for the DE to acquire a small in-plane polarized dipole moment like that of the BEs, while the BE optical activity is hardly affected by the mixing.
As the atomistic simulations show,~\cite{zielinsky2014} the DE/BE mixing has two terms, one due to hole spin flip and one due to electron spin flip. The contributions of these two terms to the optical activity of one of the DE eigenstates is constructive while to the other eigenstate it is destructive. If the two mixing terms are comparable, only one DE eigenstate will be optically active. Moreover, if one recalls that by definition the spin of a hole is opposite in direction to that of a missing electron, it follows that the symmetric BE couples to the antisymmetric DE and vice versa.

In order to be consistent with Ref.~\cite{poem2007}, we define the lowest energy state of the BE and the direction of its emission as being horizontally linearly polarized (H). From the definition of the H polarization, and the association of the $|+1\rangle,~|-1\rangle$ BE states with right- (left-) hand circularly polarized light,~\cite{benny2011prl, poem2011, kodriano2012}  it follows that this choice imposes that the lowest BE eigenstate is symmetric in the $|+1\rangle,~|-1\rangle$ basis. Therefore, if one is consistent and uses the basis $|+2\rangle,~|-2\rangle$ for describing the DE eigenstates, the optically active DE which is coupled to the symmetric BE is antisymmetric in the $|+2\rangle,~|-2\rangle$ basis. This being said, we note here that these symmetry properties are basis-dependent. We use them in the following for clarity only. Full characterization of an eigenstate is only given by associating it with a measurable property like its energy. We show below, by measuring the direction of the DE precession, that the optically active DE eigenstate is the lower energy one. This agrees also with the atomistic model calculations of Ref.~\cite{zielinsky2014}

Though only one DE eigenstate is optically active, both its eigenstates are equally connected through dipole allowed (``bright") optical transitions to the
$XX_{T\pm3}^{0}$  biexciton states, as described in Figure \ref{fig:initiate_probe}. Thus the DE as a physical two-level system (qubit) can be optically accessed and coherently controlled using these optical transitions as discussed and demonstrated below.

\subsection{The Dark Exciton - $XX^{0}_{T\pm3}$ Biexciton Optical Transitions}

\begin{figure}
\begin{center}
\includegraphics[width=\columnwidth]{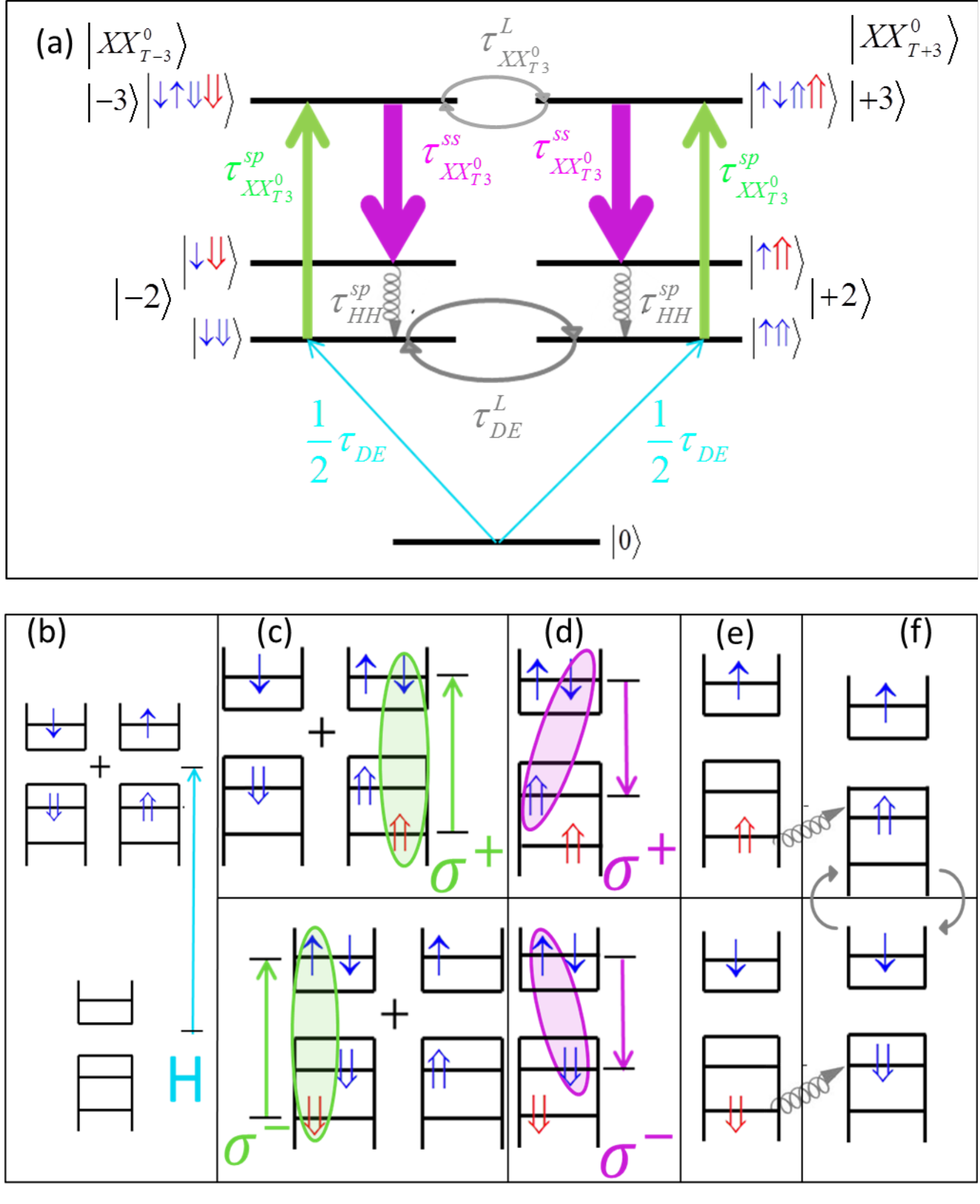}
\caption{Schematic description of the experimental steps. (a) The energy levels and optical transitions of the DE - biexciton system are shown schematically. The colors of the arrows correspond to those used in Figure \ref{fig:spectrum}. The width of the arrow represents the strength of the optical transition, with thicker arrows corresponding to stronger dipole moments.
Non-radiative transitions are inidicated by grey arrows. Curly grey downwards arrows indicate non-radiative relaxations, and the curved grey arrows represent coherent precession. The various energies and rates are listed in Table \ref{tbl:oscillator} (b) An H-polarized
resonantly tuned laser pulse excites the DE in its anti-symmetric eigenstate. (c) a left- (bottom panel ) or right- (top panel) hand circularly polarized laser pulse, resonantly tuned to the higher energy optical transition of the
$XX^{0}_{T\pm3}$ biexciton (green upright arrow in Figure \ref{fig:spectrum}) is used to probe the
state of the DE. (d) Detecting a left (bottom panel) or a right (top panel)
circularly polarized photon from the lower energy optical transition of the
$XX^{0}_{T-3}$ ($XX^{0}_{T+3}$, top) biexciton heralds the presence of a DE in a $|-2\rangle$
(lower panel) or a $|+2\rangle$ (upper panel) coherent state. (e-f) The heavy hole relaxes to the ground state via a fast spin-preserving relaxation, leaving a ground state DE in the QD. }
\label{fig:initiate_probe}
\end{center}
\end{figure}

Figure \ref{fig:initiate_probe}(a) schematically describes the states involved in the experiments and the transitions between these states. The states are represented by horizontal lines in increasing energy order. The carrier configurations, the spin wavefunction, and the total spin of each state are depicted to the side of each state. Optical transitions between these states are represented by straight arrows whose colors correspond to those used in Figures \ref{fig:spectrum}(a) and \ref{fig:initiate_probe}(b-f). The width of the arrows represents the oscillator strength of the transition: the thicker the arrow is, the stronger is the transition that it represents.
 Curly grey downward arrows represent non-radiative relaxation of the excited hole to its ground state and the curved horizontal grey arrows represent coherent Larmor precession of the DE qubit and that of the $XX^{0}_{T\pm3}$ biexciton. The rates for each transition as directly measured or independently deduced from the experiments are summarized in Table \ref{tbl:oscillator}.

\begin{table}
\begin{center}
\begin{tabular}{|c|c|c|c|}
\hline
Quantity & Symbol & Time (nsec) & Method \\ \hline \hline
BE lifetime & $\tau_{BE}$ & $0.47$ & PL, TRS \\
DE lifetime & $\tau_{DE}$ & $1.1 \times 10^3$ & PL, TRS \\
DE Larmor period & $\tau^{L}_{DE}$ & 3.1 & $g^{(2)}$~\cite{poem2010} \\
$XX^{0}_{T3}$ $sp$ radiative time & $\tau^{sp}_{XX^{0}_{T3}}$ & 6 & PL \\
$XX^{0}_{T3}$ $ss$ radiative time & $\tau^{ss}_{XX^{0}_{T3}}$ & 0.30 & TRS \\
$XX^{0}_{T3}$ Larmor period & $\tau^{L}_{XX^{0}_{T3}}$ & $>$ 5 & RECPM \\
Heavy hole relaxation time & $\tau^{sp}_{HH}$ & 0.02 &PL~\cite{poemprb2010}, TRS \\
\hline
\end{tabular}
\caption{Measured times (inverse rates)  used  in Figure \ref{fig:initiate_probe}(a). The type of measurement used to derive each quantity is listed in the rightmost column where PL stands for lines intensity ratio, TRS for pump-probe time resolved spectroscopy,  $g^{(2)}$ for intensity correlation measurements, and RECPM for resonantly excited circular polarization memory measurements.}
\label{tbl:oscillator}
\end{center}
\end{table}

In Figure \ref{fig:initiate_probe} (b)-(f) the various steps in our experiments are schematically described.
In the first step (Figure \ref{fig:initiate_probe}(b)), a resonantly tuned horizontally-linearly (H) polarized laser pulse photogenerates the DE in its lower energy, optically active, anti-symmetric eigenstate $|a\rangle$. The pulse is tuned into the vacuum-DE weak optical transition at 1.2834 eV (marked by a blue arrow in Figure \ref{fig:spectrum}(c) and Figure \ref{fig:initiate_probe} (a)).
As we show in the next section, despite the weakness of this optical transition, the intensity and the duration of the pump pulse can be tuned such that a $\pi$ pulse is obtained. In this way,  deterministic generation of the DE in a well-defined eigenstate ( $|a\rangle = 1/\sqrt{2}(|+2\rangle-|-2\rangle)$, Figure  \ref{fig:initiate_probe}(b)) is achieved.

Probing the DE presence in the QD is done by absorbtion of a second laser pulse which is resonantly tuned to the DE-$XX^{0}_{T\pm3}$ optical transition at 1.294 eV (green arrows in Figure \ref{fig:spectrum}(a) and in Figure \ref{fig:initiate_probe} (a,c)). The efficiency of the absorption is measured by the detection of light emission at 1.278 eV due to the biexciton recombination (downward magenta arrows in Figure \ref{fig:spectrum}(a) and in Figure \ref{fig:initiate_probe} (a,d)).

As shown in Figure \ref{fig:initiate_probe}(c), if a DE is present in the QD, a right (left) hand circularly
polarized $\sigma^{+}$($\sigma^{-}$) probe transfers the $|+2\rangle$
($|-2\rangle$) component of the DE population to the $XX_{T+3}^{0}$ or
$|+3\rangle$ ($XX_{T-3}^{0}$ or $|-3\rangle$) biexciton (Figure
\ref{fig:initiate_probe}(c)). Note that Pauli exclusion prevents absorption of $\sigma^{-}$ ($\sigma^{+}$) photons at this resonance energy when the DE is in the $|+2 \rangle$ ($|-2 \rangle$) spin state. Absorption of a given polarization thus directly depends on the DE state. The radiative decay of the $XX_{T+3}^{0}$
($XX_{T-3}^{0}$) biexciton then results in strong $\sigma^{+}$($\sigma^{-}$)
emission from the lower energy transition of this biexciton, which is marked
by a downward magenta arrow in Figure \ref{fig:spectrum}(a)(Figure \ref{fig:initiate_probe}(d)). The two biexcitonic states, $|+3 \rangle$ and $|-3 \rangle$ are nearly degenerate in energy, such that there is no appreciable precession between them We obtained a lower bound of 5 nsecs on the Larmor precession time of these two states (Table \ref{tbl:oscillator}). Thus, there is almost a one-to-one correspondence between the DE spin state, the polarization of the absorbed photon, and the polarization of the photon emitted from this biexciton ({\it i.e.} a $|+2 \rangle$ DE absorbs a $\sigma^{+}$-polarized photon and the corresponding biexcitonic emission is also $\sigma^{+}$-polarized, albeit at a different energy).
The emission intensity of $\sigma^{+}$($\sigma^{-}$) circularly-polarized
photons from the $XX^{0}_{T\pm3}$ biexciton is thus directly proportional to the DE  $|+2\rangle$ ($|-2\rangle$) state
population and we use this emission for probing both the overall DE population and the specific DE spin state.

While the ground state electrons in the $XX^{0}_{T\pm3}$ biexciton could recombine optically with either of the two heavy holes, the biexciton recombination is about fifty times more probable in the lower energy optical transition (Figure \ref{fig:spectrum}(a)) due to the better overlap of the ground-level wavefunctions.
This recombination leaves a DE in the QD formed by an
electron-excited-hole pair (Figure \ref{fig:initiate_probe}(e)). The hole returns to its ground level within
$\sim$20 psec via a spin-preserving phonon-mediated relaxation (Figure \ref{fig:initiate_probe}(f)).~\cite{poemprb2010, kodriano2010} Consequently, detection of a $\sigma^{+}$($\sigma^{-}$) photon from the biexciton recombination heralds the generation of a DE in a
well-defined coherent superposition of its two
eigenstates $|s\rangle$ and $|a \rangle$. The coherent DE then precesses in time\cite{poem2010} on the DE
Bloch sphere as depicted in Figure \ref{fig:experiment_scheme}(a). The
precession period of $3.09$ nsec corresponds to the energy splitting
(1.4 $\mu$eV) between the DE eigenstates, which has been previously measured.\cite{poem2010} The DE can
then be re-excited, using a circularly polarized resonant optical pulse,
thereby probing the temporal evolution of its spin state.

\subsection{Deterministic Initialization of the Dark Exciton and Measurement of its Lifetime}

We now demonstrate deterministic optical initialization of the DE.

\begin{figure}
\begin{center}
\includegraphics[width=\columnwidth]{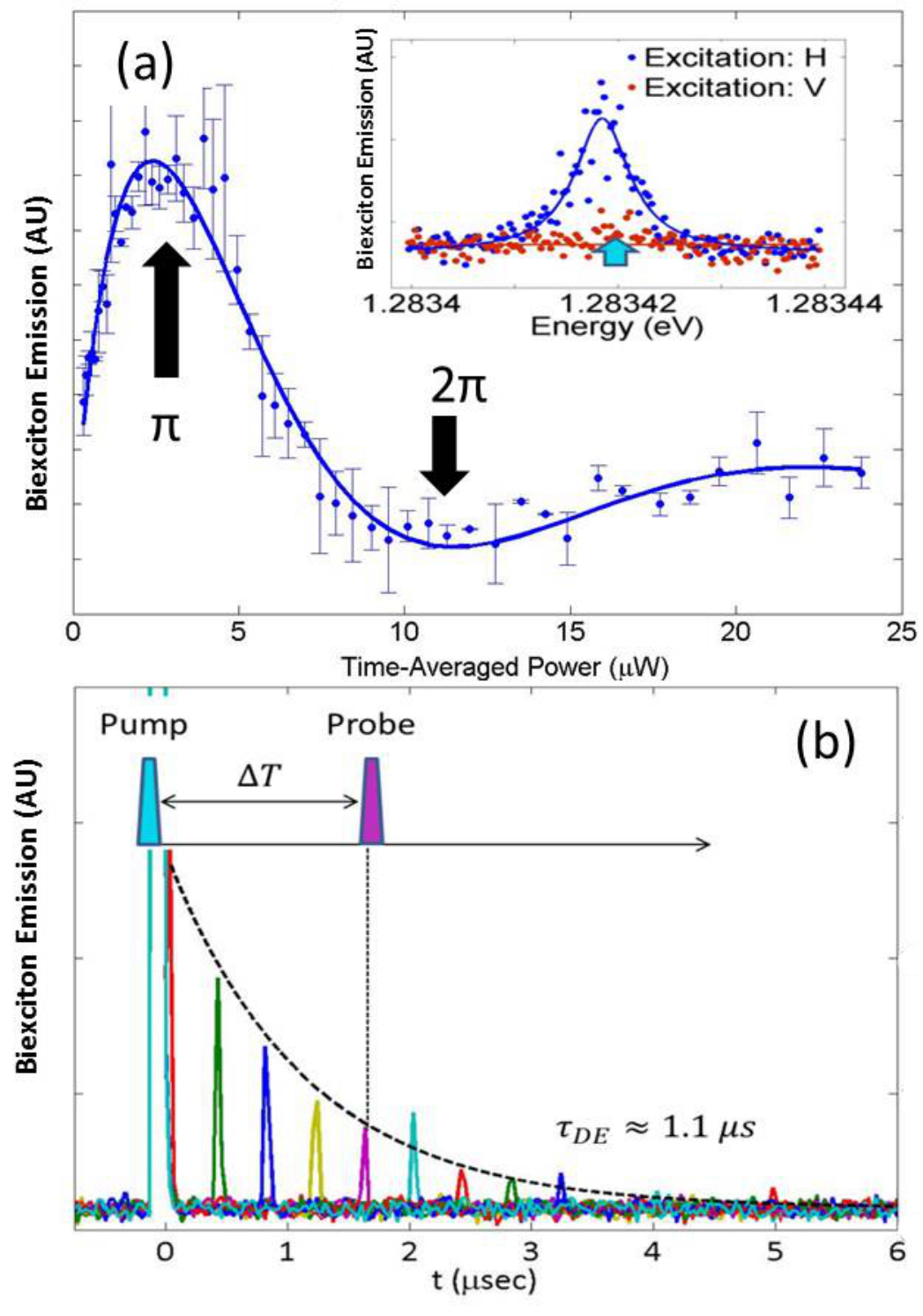}
\caption{Deterministic generation of the DE (a) Lower-energy $XX^{0}_{T\pm3}$
biexciton emission intensity (Figure \ref{fig:initiate_probe}(d)) as a function of the average
resonant excitation power into the DE absorption resonance (Figure \ref{fig:initiate_probe}(b)) with
a pulse width of 60 nsec, in the presence of a probe laser tuned to the
higher-energy $XX^{0}_{T\pm3}$  transition resonance (Figure \ref{fig:initiate_probe}(c)). The powers at
which $\pi$- and 2$\pi$-pulses are obtained are marked by vertical black
arrows. The solid line represents a model fit of the Rabi oscillations. In
the inset, the $XX^{0}_{T\pm3}$  biexciton emission intensity as a function of
the energy of the $H$ ($V$) linearly polarized excitation into the DE
optical transition is depicted by the blue (red) dots. The energy used for
resonant, deterministic excitation of the DE is indicated by a blue arrow.
(b) Lifetime of the DE as measured by a time-resolved double resonant
pump-probe experiment. The $XX^{0}_{T\pm3}$ biexciton emission is
measured as a function of the time difference between the DE excitation
$\pi$ pulse and the $XX^{0}_{T\pm3}$ probe pulse. The
inset describes the experiment sequence of pulses, where the DE is excited
at time $t=0$ (blue pulse) and its population is probed via absorption of a
time-delayed probe $\pi$-pulse (magenta). The intensity of the resulting
emitted light from the  biexciton is a measure of the magnitude of the
decaying DE population. The dashed black line is a fitted exponential decay
model with characteristic DE lifetime of 1.1 $\mu$sec.  \label{fig:rabi}}
\end{center}
\end{figure}

Figure \ref{fig:rabi}(a) displays the PL emission intensity from the
$XX^{0}_{T\pm3}$  biexciton as a function of the power and energy (inset) of
the pump laser (represented by blue arrow in Figure \ref{fig:spectrum} and Figure \ref{fig:experiment_scheme}) in the presence of the probe laser (represented by green arrow in Figure \ref{fig:spectrum} and Figure \ref{fig:experiment_scheme}). The inset was recorded
with continuous wave lasers for two rectilinear polarizations of the pump
beam. It demonstrates that the energy and polarization in which the pump
laser most efficiently generates the DE is exactly the same as that of the
DE observed in photoluminescence spectroscopy ($X_{DE}^{0}$ in Figure \ref{fig:spectrum}(b)).
The linewidth of the DE resonance ($\sim$ 5 $\mu$eV) reflects the radiative width of the transition to the biexciton, broadened by
spectral diffusion caused by the presence of the two laser beams.

In the main panel of Figure \ref{fig:rabi}(a), pulsed pump-probe measurements at a repetition rate
of 1MHz were used. Rabi oscillations are clearly observed. The pump pulse
intensities corresponding to $\pi$- and 2$\pi$-pulses are marked by vertical
black arrows. Figure \ref{fig:rabi}(a) therefore demonstrates that a proper
combination of pulse energy, polarization, duration, and power,
deterministically generates the DE.
The deterministic generation of the DE was verified in additional way.
The rate by which single photons due to recombinations of $XX_{T+3}^{0}$ biexciton are detected is in agreement with our estimations based on
the known excitation repetition rate (1MHz) and the overall light harvesting efficiency of the experimental
setup (about 1 out of  700 photons).

A comparison between the intensity and
temporal pulse width needed to obtain a $\pi$-pulse to the DE resonance with that
needed for a $\pi$-pulse to the BE directly yields the ratio between the oscillator strengths (radiative lifetimes)
of both excitons. We independently verified in this way that the DE oscillator strength
is more than three orders of magnitude weaker than that of the BE.

In Figure \ref{fig:rabi}(b), we present a direct measurement of the DE lifetime following deterministic
initialization of the DE in its $|a \rangle$ spin state. Here, the $XX^{0}_{T\pm3}$  biexciton emission is measured
for various delay times between the DE pump $\pi$-pulse and the probe pulse
to the $XX^{0}_{T\pm3}$  biexciton. The inset describes the experimental
sequence of pulses. As stated previously, the intensity of the emitted light
from the $XX^{0}_{T\pm3}$ biexciton probe is a measure of the magnitude of
the DE population. The dashed black line is a fitted exponential
decay model with a characteristic lifetime of $1.1\pm0.1$ $\mu$sec.
The fact that the measured lifetime of the DE agrees with its
measured oscillator strength establishes that the DE
decays radiatively from its singly $H$-polarized mode.

\subsection{The Coherence Time of the Dark Exciton}

The coherence time of a qubit is an important measure of its decoupling from the environment. Long coherence times are essential for potential applications in QIP. Here, we present a first experimental determination of a lower bound for the coherence time of the DE.

When the DE spin state is a coherent superposition of its two eigenstates $|s\rangle$ and $|a\rangle$, the phase between the two eigenstates varies periodically with a period time (3.09 nsec~\cite{poem2010}) given by the Planck constant divided by the energy difference between the eigenstates (1.4 $\mu$eV).  This periodic behavior is commonly described as a precession along the equator of the qubit's Bloch sphere~\cite{kodriano2012} (Figure \ref{fig:experiment_scheme}(a)). The qubit coherence time can be loosely defined as the characteristic decay time for this precession.

As shown in Figure \ref{fig:initiate_probe}(c), a $\sigma^{+}$($\sigma^{-}$) probe photon transfers the $|+2\rangle$ ($|-2\rangle$) component of the DE population to the $XX_{T+3}^{0}$
($XX_{T-3}^{0}$) biexciton. Absorption of a photon of given polarization thus directly depends on the DE spin state. The $XX_{T+3}^{0}$ ($XX_{T-3}^{0}$) biexciton radiatively decays within 300 psec resulting in emission of a $\sigma^{+}$($\sigma^{-}$)
polarized photon, which reveals the state of the DE when the probe photon was absorbed.
Thus, circular polarization-sensitive intensity autocorrelation measurements of the
the $XX_{T\pm 3}^{0}$ spectral line in the presence of the probe laser provide a direct measurement of the temporal evolution of the DE spin state. The first detected $\sigma^{+}$-polarized photon heralds the DE presence in the $|+2\rangle$ coherent state, while the second detected $\sigma^{+}$($\sigma^{-}$) polarized photon projects the DE on the $|+2\rangle$ ($|-2\rangle$) state at the time when the probe photon was absorbed.

\begin{figure}[tb]
\begin{center}
\includegraphics[width=0.89\columnwidth]{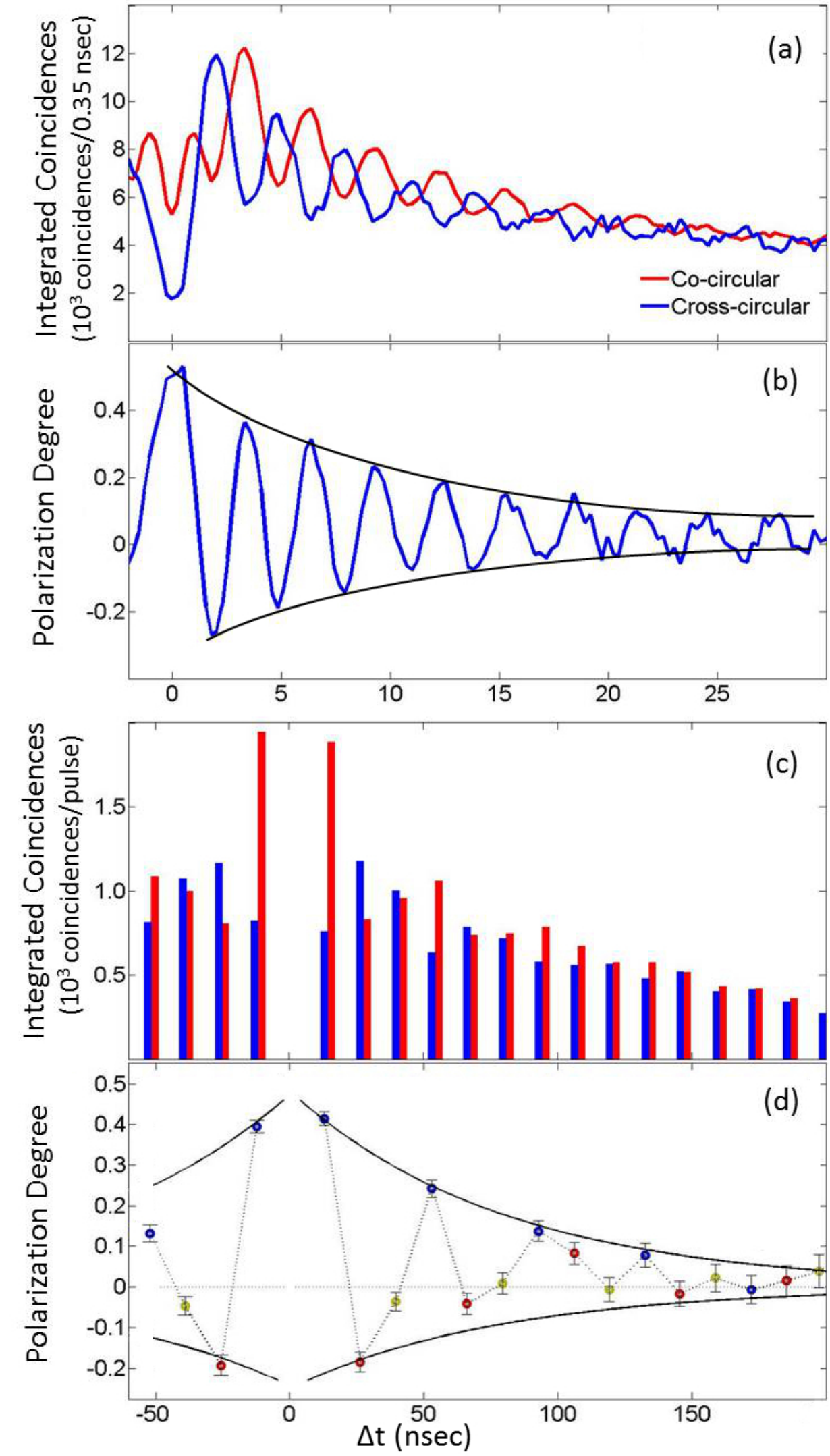}
\caption{(a) Polarization-sensitive intensity autocorrelation measurements of the
$XX_{T\pm3}^{0}$ biexciton (vertical-magenta arrow in Figure \ref{fig:spectrum}) under resonant cw excitation (vertical green arrow in Figure \ref{fig:spectrum}). The red (blue) lines display the co- (cross-) circular coincidence measurements. (b) The degree of circular polarization of the second detected photon as a function of its detection time as deduced from the measurements in (a). The decay of the polarization oscillations is overlaid by an exponential decay curve (solid black line) with characteristic time of $\sim$ 25 nsec. (c) Similar to a) but under
pulsed excitation at a repetition rate of 76 MHz. The red (blue) bars
display the integrated number of coincidences emitted after each laser pulse
for co- (cross-) circularly polarized photons. (d) Like (b) but for the pulsed excitation presented in c). From
the temporal decay of the amplitude of the oscillations (solid black lines), a
lower bound of 100$\pm$20 nsec for the DE coherence time is deduced. \label{fig:coherence}
}
\end{center}
\end{figure}

In Figure \ref{fig:coherence}(a), black (red) solid lines present intensity autocorrelation measurements of the  $XX_{T\pm 3}^{0}$ spectral line under resonant cw excitation for co-circularly (cross-circularly) polarized photon pairs. By subtracting the cross-circular measurement from the co-circular measurement and normalizing by their sum, the temporal evolution of the degree of circular polarization of the second photon is obtained (Figure \ref{fig:coherence}(b)). The observed periodic oscillations in the degree of circular polarization result from the coherent precession of the DE spin state. We note that the maximal degree of polarization obtained is approximately 0.5. This is expected, considering the limited temporal resolution of our detectors.~\cite{poem2010} In addition, we note that the oscillations decay with a characteristic decay time of about 25 nsec. This polarization decay is a lower bound on the DE coherence time, since no polarization oscillations can be observed beyond the DE coherence time.
Under cw probe excitation, the polarization oscillations also decay for another reason. Under these conditions, the probability that the detected second photon results from the absorption of only one probe photon, immediately after the photon which heralds the DE, decreases quickly with time. The absorption of each photon followed by spontaneous biexciton recombination, which is not detected, reduces the measured polarization oscillations. Indeed, we observe that the cw polarization decay time strongly depends on the probe light intensity (not shown).

In Figure \ref{fig:coherence}(c), we present a similar measurement using pulsed probe light at a repetition rate of approximately 76MHz. Red (blue) bars display the integrated number of co- (cross-) circularly polarized coincidences after each laser pulse.  The measured degree of circular polarization of the second photon as a function of the pulse time is given in Figure \ref{fig:coherence}(d).
Since the DE period is about 3 nsec while the laser period is about 13 nsec, the periodicity of the measured polarization is expected to be about 3$\times$13 nsec, or about 3 laser pulses. This is exactly what we observe in
Figure \ref{fig:coherence}(d).
The maxima (minima) of the oscillations are represented with blue (red) markers and occur every 3 pulses. The polarization degree decays with a characteristic time of 100$\pm$20 nsec. This measurement, therefore, sets a lower bound on the
coherence time $T_{2}^{*}$ of the DE. The actual time is probably longer, since the repeated
pulsed excitation shortens the polarization decay time.

\subsection{Control of the DE Spin State using a Picosecond Optical Pulse}

After demonstrating deterministic photogeneration of the DE in a well defined pure state, we present here an experimental demonstration of coherent control (``rotation") of the DE state using a short optical pulse. We rotate the DE state from the eignestate in which it was photogenerated to a coherent superposition of its two eigenstates using a $\sim$ 10 psec pulse, which is more than five orders of magnitude shorter than the measured lifetime of the DE and at least four orders of magnitude shorter than the DE coherence time.

\begin{figure}
\begin{center}
\includegraphics[width=0.9\columnwidth]{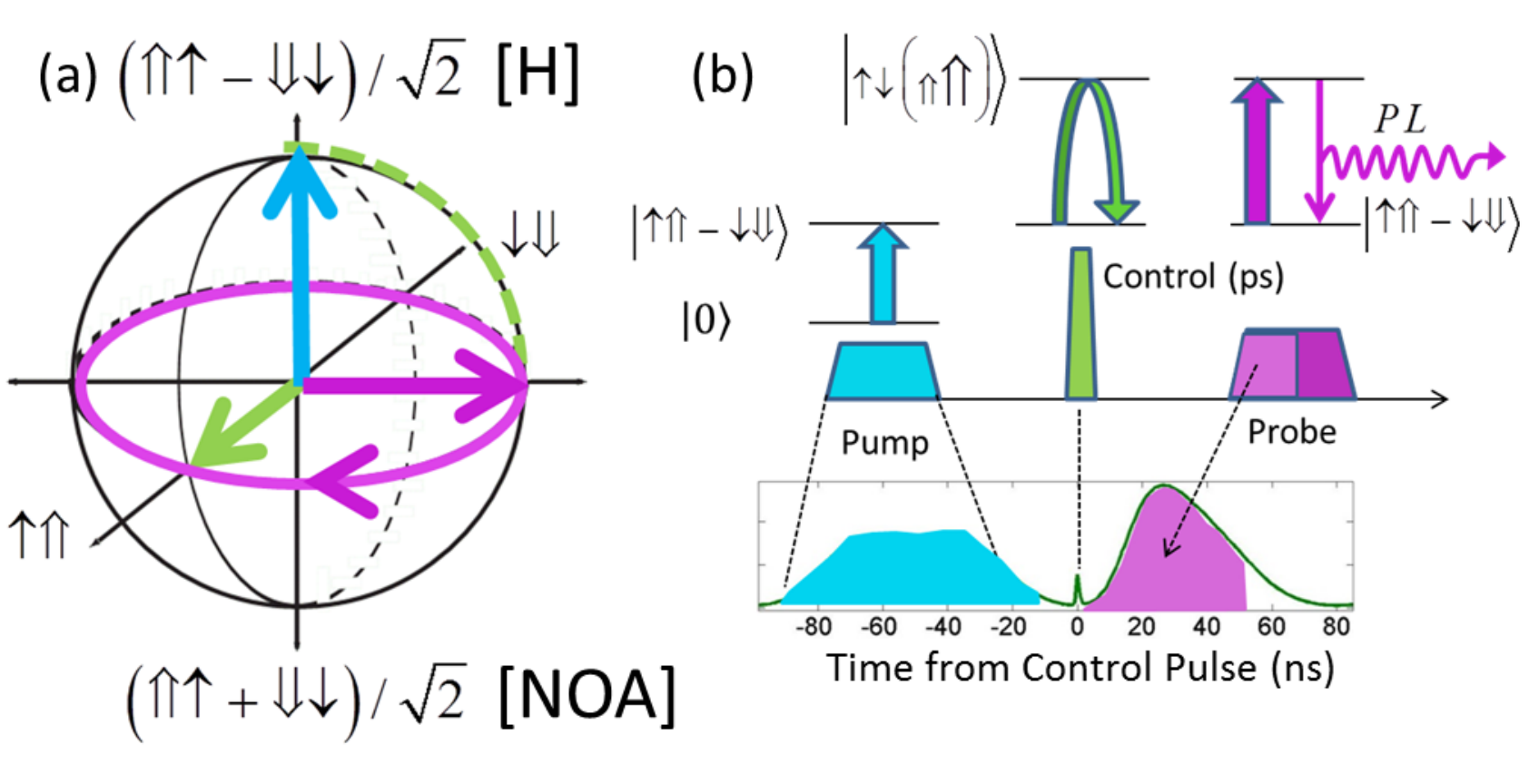}
\caption{
(a) Schematic description of the precession of the DE on the Bloch sphere.
The pump $\pi$-pulse (blue arrow, Figure \ref{fig:initiate_probe}(b))
deterministically generates the DE in its lower-energy eigenstate (north pole
of the sphere). The positively detuned circularly polarized control 2$\pi$-pulse (green
arrow, Figure \ref{fig:initiate_probe}(c)) ``rotates" the DE by an angle of
$\pi$/2 about the polarization direction (green arrow) and brings the DE
within the pulse duration (few psecs) to the equatorial plane of the Bloch
sphere (dashed green line). The DE then precesses clockwise in time as described by
the magenta trajectory on the Bloch sphere. The optical activity of the DE eigenstates ([H] for H-polarized and [NOA] for no optical activity) is shown to the right of the spin configuration. (b) Schematic description of the
pulse sequence used in this experiment. A resonant DE $\pi$-pulse (blue
arrow, Figure \ref{fig:initiate_probe}(b)) is used to deterministically excite the DE in the [H] eigenstate.
Then a properly detuned from the biexciton resonance, circularly polarized, 2$\pi$,
picosecond pulse is used to rotate the DE to its $|+2\rangle$ coherent spin
state. A long polarized probe pulse (magenta) is then used to probe the DE
precession, which gives rise to temporal oscillations in the circular
polarization of the emitted light from the biexciton transition (Figure
\ref{fig:initiate_probe}(d)). The actual measured temporal sequence of
pulses is shown below the schematic description.
\label{fig:experiment_scheme}}
\end{center}
\end{figure}

Figure \ref{fig:experiment_scheme} presents schematically (a) the rotation of the DE spin state on the Bloch sphere during this experiment and (b) the sequence of pulses used to experimentally demonstrate the rotation. The pump pulse deterministically writes the DE in its lower energy eigenstate, represented by the north pole of the Bloch sphere in Figure \ref{fig:experiment_scheme}(a). An eigenstate does not evolve in time, and the DE remains in this eigenstate as it radiatively decays.
For use as a qubit, it must be possible to control the spin state of the DE and to rotate this state at will to any point on the Bloch sphere.
We do this similarly to the way by which we\cite{poem2011, kodriano2012} and recently Muller {\it et al.}\cite{muller2013} controlled the BE spin state.
A detuned 2$\pi$-area optical pulse is used to transfer the DE population through the $XX^{0}_{T\pm3}$ resonance and back into the DE. During this transfer, the
DE state acquires a relative phase difference between its two eigenstates which is dependent on the detuning from resonance of the optical
pulse.\cite{economou2007} This phase accumulation is described on the Bloch
sphere (Figure \ref{fig:experiment_scheme}(a)), as a rotation around the
direction of the pulse circular polarization.\cite{kodriano2012, muller2013}
No detuning results in a $\pi$ rotation, while negative (positive) detuning
results in larger (smaller) rotation angles.\cite{kodriano2012, muller2013,
economou2007}
Similarly to the separate carriers,\cite{ramsay2008, greilich2011,
degreve2011, godden2012} full DE control in this manner requires two optical pulses.

Figure \ref{fig:experiment_scheme}(b) shows the pulse sequence used in this
experiment. After deterministic generation of the DE using a 60 nsec H-polarized pulse, a
very short ($\sim$10 psec) 2$\pi$ area $\sigma^{+}$-polarized control pulse,
positively (negatively) detuned from the $XX^{0}_{T\pm3}$ resonance, is applied to rotate the DE
state $\pi/2$ ($3\pi/2$) radians around the right hand circular polarization direction on the Bloch sphere. A 60 nsec right hand circularly-polarized cw probe is then used to re-excite the
DE to the $XX^{0}_{T\pm3}$ biexciton for measuring the time evolution of the DE
spin state during these subsequent 60 nsec. The resulting $XX^{0}_{T\pm3}$
emission is detected after a circular polarizer which matches the circular
polarization of the probe pulse. By subtracting and dividing by the sum of
the time dependent signals from the  cross- and co-circular
polarizations, the degree of circular polarization during the probe pulse is
obtained. The entire sequence of pulses lasts $\sim$120 nsec and the
repetition rate is 1MHz. In Figure \ref{fig:experiment_scheme}(b), the
measured pulse sequence is shown below the schematic description.

\begin{figure}
\begin{center}
\includegraphics[width=\columnwidth]{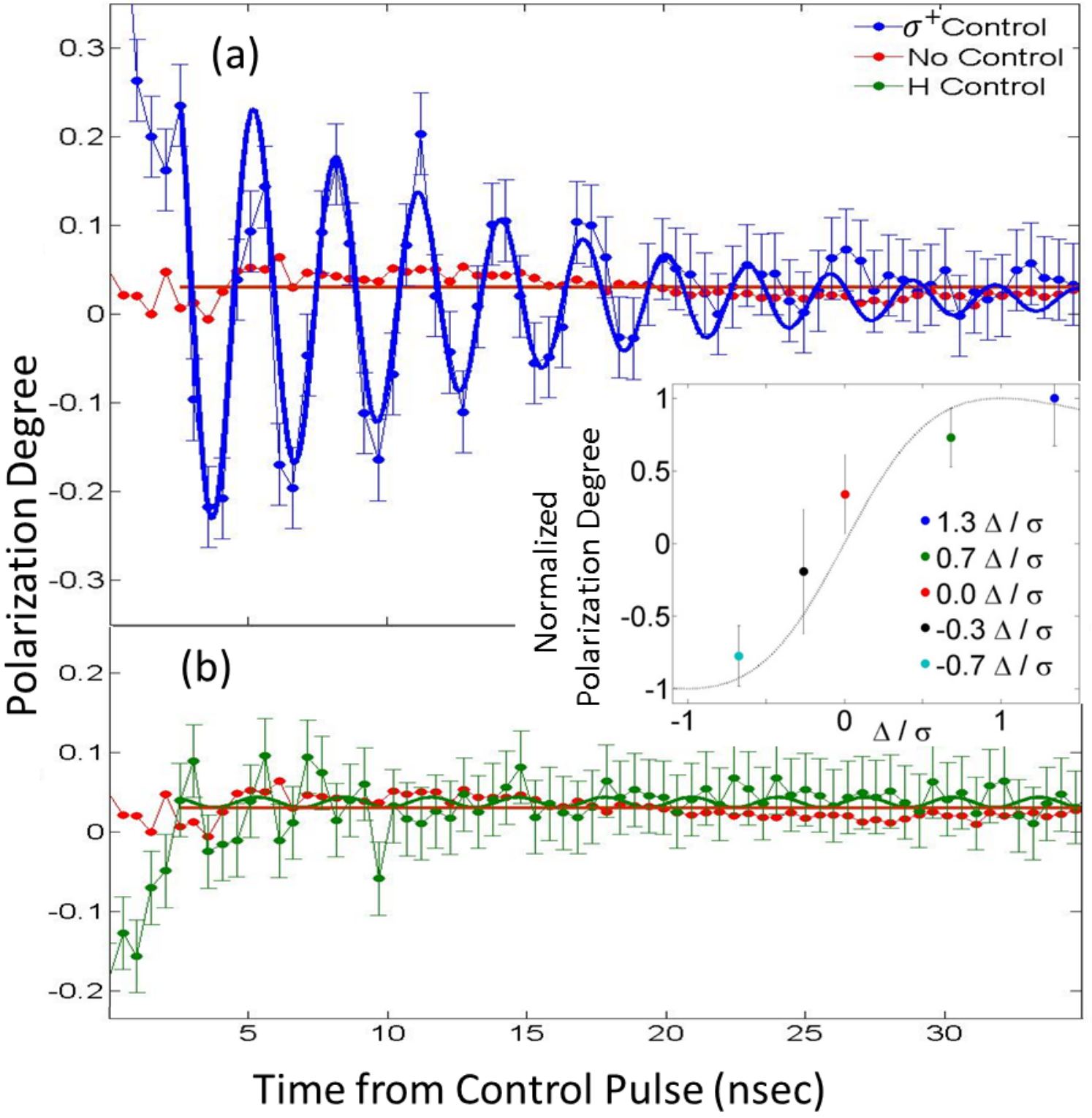}
\caption{Experimental demonstration of the DE spin control (a) [(b)] The
measured (points) degree of circular polarization of the $XX^{0}_{T\pm3}$
emission as a function of time after the application of a $\sigma^{+}$- (right hand
circularly)- [$H$- (horizontal linearly )] polarized control pulse (blue)
[(green)], compared to that measured in the absence of a control pulse
(red). The solid line in (a) represents a decaying sinusoidal model fit.
Note that while no polarization oscillations are observed for the linearly
polarized control pulse, they are clearly visible for the circularly
polarized one. Points (line) in the inset show the measured (model)
dependence of the normalized oscillation amplitude on the detuning.
\label{fig:data}}
\end{center}
\end{figure}

In Figure \ref{fig:data}(a), we present the measured (points) degree of
circular polarization as a function of time after the control pulse
during the first 35 nsec of the
probe pulse for a detuning energy of $\Delta/\sigma \approx 0.7$ for a
$\sigma^{+}$-polarized control pulse. Here, $\Delta$ is the detuning from
resonance and $\sigma = 100$ $\mu$eV is the full spectral width of the
control laser at half maximum. The overlaid solid line presents a best-fit
model of an exponentially-decaying sinusoidal function $f(t) =
P_{0}\sin(2\pi t/\tau^{L}_{DE}) \exp[-t/T_{PD}]$, where $P_{0}$ is the initial
polarization degree, $\tau^{L}_{DE}$ $\sim 3$ nsec is the DE precession time (Table \ref{tbl:oscillator}) and
$T_{PD}$ is a characteristic polarization decay time, which under these
conditions of strong cw re-excitation it is relatively short ( only $\sim 20$ as opposed to $\sim 100$ nsec
under pulsed excitation as in
Figure \ref{fig:coherence}(d)).

In the inset to Figure \ref{fig:data}, points present the actual values of the
best-fitted $P_{0}$ (normalized to 1 at maximum) as a function of the
detuning energy. The solid line presents the theoretically expected
dependence $P_{0}(\Delta) = \sin(\pi - 2
\arctan[\Delta/\sigma])$.\cite{economou2007} The maximal degree of initial
polarization achieved in these experiments at a detuning of $\Delta/\sigma
\approx 1.0$ is about 30\%. Two factors limit the measured initial
polarization in these experiments. The first is the temporal resolution of
our detectors, which limits $|P_{0}|<0.6.$ as discussed in Ref. \cite{poem2010}
The second factor is the repetition rate at which the experiments are
conducted. The pulse-to-pulse separation is comparable to the DE lifetime.
Therefore, in $\sim$40\% of the cases, the residual DE population prevents
the absorption of the pump pulse, thus further reducing $|P_{0}|$. A pulse
sequence in which the QD is emptied after the probe pulse and before the DE
is photogenerated again, can in principle yield closer to unity fidelity.

We note in Figure \ref{fig:data}(a) that for a positive (negative, see inset) detuning there is a delay of a quarter (three quarters) of a precession period  between the control pulse time and the time at which the maximal polarization degree is obtained.
This is an unambiguous experimental verification for the clockwise precession of the DE qubit (Figure \ref{fig:experiment_scheme}(a)). This establishes experimentally, that the optically active DE eigenstate, into which we deterministically initiate the DE, is the DE lowest energy eigenstate.

For completion, we present in Figure \ref{fig:data}(b), the measured degree of
circular polarization at the same detuning as in Figure \ref{fig:data}(a) but
this time the control pulse is linearly $H$-polarized. Clearly, as expected,
no polarization oscillations are observed in this case, similar to the case
with no control pulse at all (red line).

\section{Conclusions}

In conclusion, we demonstrated that the long-lived, quantum dot-confined dark exciton can be
deterministically generated in a pure and well-defined spin eigenstate using
a single optical pulse. We also showed that its spin state can be coherently controlled using
single picosecond pulse, resonantly tuned to or slightly detuned from a biexciton state.
In addition, we obtained lower bounds of about 1 and 0.1 $\mu$sec on the life and coherence time of the dark exciton,
respectively.

These demonstrations suggest that the dark exciton may have possible advantages over its constituents, the
separate charge carrier as QD confined spin qubit:
(a) The DE qubit requires no external magnetic field. However, vertically applied magnetic field can be easily applied to the
DE qubit thereby increasing or tuning its Larmor frequency.
(b)      The DE coherence time as measured without spin echo is at least an
order of magnitude longer than that of its constituents, under similar
conditions.
(c)      The DE can be initialized deterministically by one single optical
pulse, unlike the separate carriers, which require optical pumping,
containing at least sequences of few pulses.
(d)      The DE can be reset (i.e the QD be emptied from both carriers and be
prepared for new initialization) by one single voltage
pulse,\cite{mcfarlane2009} which can be applied to all the QDs in the
sample. Resetting the separate carrier qubits requires optical pumping,
which may differ from one QD to another, and in principle, requires more
time.
(e)      The optical control of the DE is performed through a biexcitonic
state. This state has two non-degenerate optical transitions,
facilitating resonant detection and resonant excitation at two different
energies, and preventing blinding the detectors by the exciting laser field.
This is in clear contrast with the separate carriers, which
require one resonant excitation into a charged exciton (trion) state.

There are clear similarities between the DE qubit and the
separate carrier qubits, thus, operations that were demonstrated or proposed
with charge qubits are likewise possible with the DE qubit. These include, in particular,
entangling the DE spin with that of the emitted biexciton photon \cite{akopian2006, gao2012, degreve2012, shaibley2013} and
coupling one QD confined DE qubit with another DE qubit, either by spin-spin
interactions,~\cite{bayer2001, gywat2002} or by microcavity photons.~\cite {imamoglu1999, press2007, hu2008}

These features and increased flexibility in the initialization, possible
reset and control of the dark exciton spin qubit, make it a semiconductor based solid state
spin qubit with great potential for basic scientific studies and novel
applications.

\begin{acknowledgments}
The support of the Israeli Science Foundation (ISF), the Israeli Ministry of
Science and Technology (MOST), and
that of the Technion's RBNI are gratefully acknowledged.
\end{acknowledgments}

\bibliography{optical_control_prl_bib}

\end{document}